\documentclass[%
 aip,
 amsmath,amssymb,
 reprint,%
]{revtex4-1}

\usepackage{graphicx}
\usepackage{dcolumn}
\usepackage{bm}

\usepackage[utf8]{inputenc}
\usepackage[T1]{fontenc}
\usepackage{mathptmx}
\usepackage{etoolbox}
\usepackage{array}

\makeatletter
\def\@email#1#2{%
 \endgroup
 \patchcmd{\titleblock@produce}
  {\frontmatter@RRAPformat}
  {\frontmatter@RRAPformat{\produce@RRAP{*#1\href{mailto:#2}{#2}}}\frontmatter@RRAPformat}
  {}{}
}%
\makeatother
\begin{document}

\preprint{AIP/123-QED}

\title[Binary Colloidal Networks]{Shear-Induced Structural Convergence but Formation-History-Dependent Yielding in Sequentially Gelled Binary Colloidal Networks}
\author{Alexander Kaltashov}
\author{Safa Jamali}%
 \email{s.jamali@northeastern.edu.}
\affiliation{Department of Mechanical and Industrial Engineering, Northeastern University, Boston, MA, USA 02131}%
\affiliation{Department of Chemical Engineering, Northeastern University, Boston, MA, USA 02131}%

\date{\today}

\begin{abstract}
Multicomponent colloidal gels can exhibit mechanical responses that depend not only on interaction strengths but also on the temporal pathway by which their networks form. Here, we use particle-based simulations to investigate the steady-shear deformation of binary colloidal gels assembled by sequential gelation with tunable delay time and dominant interspecies attractions. Although varying the gelation delay produces markedly different quiescent morphologies, ranging from well-mixed networks to coarse shell-core structures, steady shear drives the systems toward structurally convergent, mixed states as quantified by cluster, connected-component, and coordination analyses. This structural convergence, however, does not imply rheological equivalence. The transient stress response remains strongly dependent on gelation delay and interspecies attraction strength. For moderate interspecies attractions, increasing delay enhances the stress overshoot, particularly at high shear rates. For stronger interspecies attractions, initially heterogeneous gels exhibit two-step yielding at low shear rates, indicating distinct deformation and restructuring processes. These results show that sequential gelation can imprint a persistent rheological memory in binary colloidal gels, even when shear substantially erases differences in common structural descriptors.
\end{abstract}

\maketitle

\section{\label{sec:level1}Introduction}

Multicomponent gel systems have been shown to exhibit remarkable, non-intuitive mechanical properties. A seminal work by Gong et al. \cite{Gong2003} showed that combining a rigid, brittle polymer network with a flexible one can create a double-network hydrogel whose toughness far exceeds than the sum of its constituents. This discovery highlighted the critical role of network heterogeneity and sacrificial deformation mechanisms, and spurred extensive interest in the exploration of multicomponent polymer hydrogels\cite{Zhu2024,Sun2012,Zhao2014}. In a similar vein, multicomponent colloidal gel systems have attracted increasing interest as well, with recent work exploring how varying inter- and intraspecies interactions can tune both structure and rheology. For example, numerical studies of colloidal double and triple networks have shown that the interplay of different gel networks in multicomponent colloidal gels profoundly affects mechanical responses\cite{FerreiroCrdova2020}, while studies of binary gels demonstrate that tuning interspecies attractions leads to distinct architectures and rheological behavior reminiscent of polymer double networks \cite{Mugnai2025,MaciasRodriguez2024}.

In colloidal gels, however, mechanical response is not determined solely by the equilibrium or quiescent structure\cite{Colombo2014,Johnson2018}. Because these materials are arrested, disordered networks formed through kinetic pathways\cite{Haghighi2025,Fenton2023}, their rheology is strongly influenced by formation history, bond lifetime, mesoscale heterogeneity, and the manner in which force-bearing structures rearrange under deformation\cite{Landrum2016,Nabizadeh2022,Das2022,Countryman2025,Vinutha2023,Rocklin2021}. Under shear, colloidal gels may elastically deform, coarsen, fragment, densify, mix, or fluidize depending on the competition between imposed deformation, Brownian relaxation, and interparticle attraction strength\cite{Moghimi2021,Koumakis2013,Koumakis2015,Hsiao2014,Hsiao2012}. As a result, the transient stress response—particularly the stress overshoot, yielding strain, post-yield relaxation, and possible multi-step yielding—can contain information about the deformation pathway that is not obvious from static structural metrics alone\cite{Boromand2017,Jamali2017,Koumakis2011,Moghimi2017}. This distinction is especially important for multicomponent colloidal gels, where different species may form domains, interpenetrating networks, or mixed composite structures depending on their interaction strengths and assembly sequence\cite{Mugnai2025,FerreiroCrdova2020}. In such systems, applied shear can play two competing roles. On one hand, deformation can erase aspects of the initial morphology by breaking, rearranging, and re-forming particle contacts, potentially driving gels with different quiescent architectures toward similar flowing or partially restructured states. On the other hand, the stress response may retain a memory of the initial network architecture if different formation pathways create distinct load-bearing backbones, domain sizes, or populations of weak and strong bonds\cite{Zaccone2009,Smith2024,Mangal2023}. Whether shear-induced structural convergence also implies rheological equivalence therefore remains an open question.

While tuning inter- and intra-species interactions offers one route of controlling gel architecture and rheology, it is not the only tuning mechanism available in the design of multicomponent colloidal systems. Indeed, exploiting the plethora of directed self-assembly mechanisms available to colloids \cite{Grzelczak2010,Guo2018} enables temporal control as an additional avenue for gel design. By exploiting the frequently independent gelation mechanisms of a system’s individual components through changes in solvent conditions such as pH, temperature or ionic strength, it is possible to induce sequential, rather than concurrent, gelation of the constituents. Although this sequential gelation strategy is still relatively underexplored in the context of colloidal gels, it can provide a powerful means to design gels with tunable and drastically different gel morphologies from the same set of components \cite{Immink2019,DiMichele2013,Appel2015,Kaltashov2026}.

Previously, we studied the quiescent structural properties of binary colloidal gels constructed through sequential gelation with a tunable delay \cite{Kaltashov2026}. Specifically, we showed that in binary systems where attractive interspecies interactions dominate over intraspecies interactions, varying the duration of the sequential gelation delay enables systematic control over species mixing, structural feature size, and the emergence of mixed versus shell-core-like morphologies. However, the rheological consequences of this pathway-dependent structure remain unresolved. In particular, it is not known whether the distinct quiescent architectures produced by sequential gelation remain mechanically distinguishable once the systems are subjected to steady shear, or whether strong interspecies attractions drive the materials toward a common structural and rheological state.

Here, we address this question by numerically investigating the steady-shear deformation of sequentially gelled binary colloidal networks with weak intraspecies attractions and dominant interspecies attractions. We examine how the gelation delay and interspecies attraction strength jointly control shear-induced structural evolution and transient stress response. We find that, despite pronounced differences in quiescent morphology, the networks undergo shear-induced structural convergence according to several connectivity-based descriptors, including cluster size, single-species largest connected components, and species-specific coordination numbers. However, this structural convergence does not imply rheological equivalence. The stress response remains strongly dependent on both formation history and interspecies attraction strength, with delayed gelation enhancing stress overshoots at high shear rates and strong interspecies attractions producing two-step yielding at low deformation rates. These results demonstrate that sequential gelation can imprint a persistent rheological memory in multicomponent colloidal gels, even when steady shear substantially reduces differences in common structural descriptors.

\section{Methods}

\subsection{Quiescent Structure Construction}

\begin{figure}[]
    \centering
    \includegraphics[width=\linewidth]{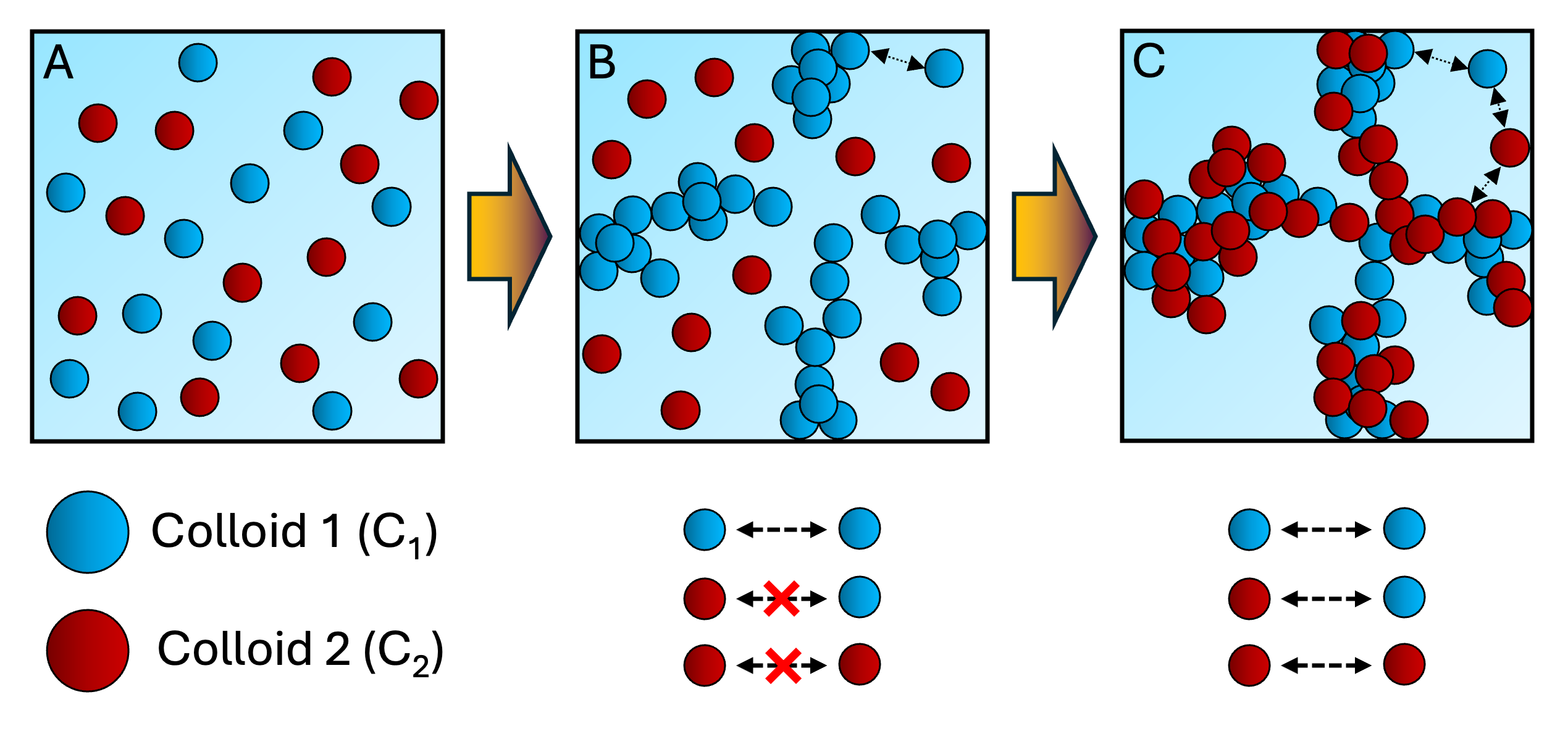}
    \caption{The studied sequential gelation process. \textbf{[A]} $C_1$ and $C_2$ particles are randomly dispersed in solvent. \textbf{[B]} $C_1$ particles begin interacting with each other and form an initial structure. \textbf{[C]} After a delay, $C_2$ particles begin interacting with $C_1$ and each other, and are incorporated into the final quiescent structure.}
    \label{fig:Fig1}
\end{figure}

All numerical simulations were performed using a modified version of the HOOMD-blue package for particle system simulation. As described previously\cite{Kaltashov2026}, we simulated the sequential self-assembly of a 20\% by volume binary mixture of N = 442,206 colloidal particles split evenly into colloidal species Colloid 1 ($C_1$) and Colloid 2 ($C_2$). Initially, the solvent conditions were chosen such that only $C_1$ particles experienced attractive interactions, thereby allowing for the formation of $C_1$ particles into an initial structure without interference from the $C_2$ population. At time $t_{delay}$ (representing the sequential gelation delay), the solvent conditions were changed to enable the interaction of $C_2$ particles with $C_1$ and each other, allowing for the incorporation of the $C_2$ population into the resulting gel structure. This sequential process is visually illustrated in Figure \ref{fig:Fig1}.

Self-assembly was modeled for a total of 5,500$\tau_D$ (diffusive time $\tau_D = (6\pi\eta a^3)(k_BT)^{-1}$, where $\eta$ is solvent viscosity, $a$ is mean colloidal hydrodynamic radius, $k_B$ is the Boltzmann constant, and $T$ is temperature). The first 200$\tau_D$ of this process was modeled using Dissipative Particle Dynamics (DPD)\cite{Boromand2017,Nabizadeh2021,Mangal2024},
and the remaining 5,300$\tau_D$ was modeled using Langevin Dynamics (LD) (Supplementary Material, Section S1). This choice of model was specifically made to ensure that during the initial stages of gelation, the essential hydrodynamic effects\cite{Ness2017, Varga2016, Furukawa2010} are preserved through DPD's explicit treatment of fluid particles in the background. The gel structures observed at 5,500$\tau_D$ were then subjected to steady shear deformation, as described in the next section. Structures were constructed with sequential gelation delay lengths of $t_{delay}$ = 0 (non-sequential gelation), 20$\tau_D$, 100$\tau_D$, 200$\tau_D$ and 3,000$\tau_D$.

Aside from representing different colloidal species, the populations of $C_1$ and $C_2$ particles are otherwise identical; both possess the same density and mean hydrodynamic radius of $a = 1.0 \pm  0.05$ (a size polydispersity was included in order to suppress unrealistic crystallization that arises in models neglecting hydrodynamic interactions\cite{Zia2014,Pusey2009,Zaccarelli2009,Henderson1996}). Interparticle interactions were modelled using the Morse Potential form:

\begin{equation}
U(h_{ij})=D_0(e^{-2\kappa h_{ij}}-2e^{\kappa h_{ij}})
\end{equation}

Where $D_0$ represents the interaction well depth, $\kappa^{-1}$ represents attraction range (in all cases, short range of $\kappa = 30$ was used) and $h_{ij}$ describes the interparticle surface-to-surface distance. Three distinct Morse Potential interactions are considered in this study; weak intraspecies interactions between $C_1$ particles ($D_{0(1-1)} = 5k_BT$), weak intraspecies interactions between $C_2$ particles ($D_{0(2-2)} = 5k_BT$), and strong interspecies interactions between $C_1$ and $C_2$ particles (described by $D_{0(1-2)} = 10k_BT$ or $20k_BT$). 

\subsection{Steady Shear Deformation}

After self-assembling and aging for 5,500$\tau_D$, the binary gel structures were subjected to steady shear deformation for a total of 10 strain units ($\gamma$), with the system’s thermal energy and interparticle interactions being consistent with those used in gel formation. Shear deformation conditions (with $\dot{\gamma}$ denoting the applied deformation rate) are described both in terms of Peclet number ($Pe = (6\pi\eta\dot{\gamma}a^3)(k_BT)^{-1}$), the ratio comparing shear-induced and Brownian motion, and Mason number ($Mn = (6\pi\eta\dot{\gamma}a^3)D_0^{-1}$), the ratio between shear force and the system’s characteristic attractive forces. Because each system contains several different interparticle interactions of different strengths, the one with the largest $D_0$ was selected as the characteristic attractive force for the purposes of reporting $Mn$.

Steady shear simulations were performed using Brownian Dynamics, via the following equation of motion:

\begin{equation}
\mathbf{r}_i(t+\Delta t) = \mathbf{r}_i(t)+\frac{\Delta t}{6\pi \eta a}(\mathbf{F}_i^C+\mathbf{F}_i^B)+\mathbf{v}_{flow}(\mathbf{r}_i)\Delta t
\end{equation}

Where $\mathbf{r}_i$ is position of particle $i$, $\Delta t$ is integration timestep size, $\eta$ is solvent viscosity, $\mathbf{F}_i^C$ is the total attractive force on particle $i$ (modeled via Morse Potential interactions), $\mathbf{F}_i^B$ is the random Brownian force acting on particle $i$, and $\mathbf{v}_{flow}$ is the background velocity field of the fluid, (defined as $\mathbf{v}_{flow}(\mathbf{r}_i) = \dot{\gamma} y _i\hat{\mathbf{e}}_x$). 

\section{Results}

\subsection{Quiescent Gel Structures}

The initial gel structure and morphology will have a significant effect on the yielding behavior\cite{Bantawa2023,Mangal2023}. Hence, renditions of all studied quiescent structures are displayed in Figure \ref{fig:Fig2}A. Systems with $D_{0(1-2)} = 10k_BT$ and $D_{0(1-2)} = 20k_BT$ both exhibited the same trends with increasing $t_{delay}$. As $t_{delay}$ was increased, the resulting morphologies transitioned from well-mixed networks to segregated shell-and-core structures. Larger values of $t_{delay}$ also produced gels with progressively larger structural features, including larger pores and a lower overall surface area. A detailed description of how variations in $t_{delay}$ influence structure in systems with $D_{0(1-2)} = 10k_BT$ is presented in our earlier work\cite{Kaltashov2026}. Qualitatively, systems with $D_{0(1-2)} = 20k_BT$ exhibit similar morphologies to their counterparts with $D_{0(1-2)} = 10k_BT$ constructed with the same $t_{delay}$, albeit with a slightly higher surface area (Supplementary Material, Figure S1).

\subsection{Infomap Cluster Analysis}

\begin{figure*}[t]
    \centering
    \includegraphics[width=\textwidth]{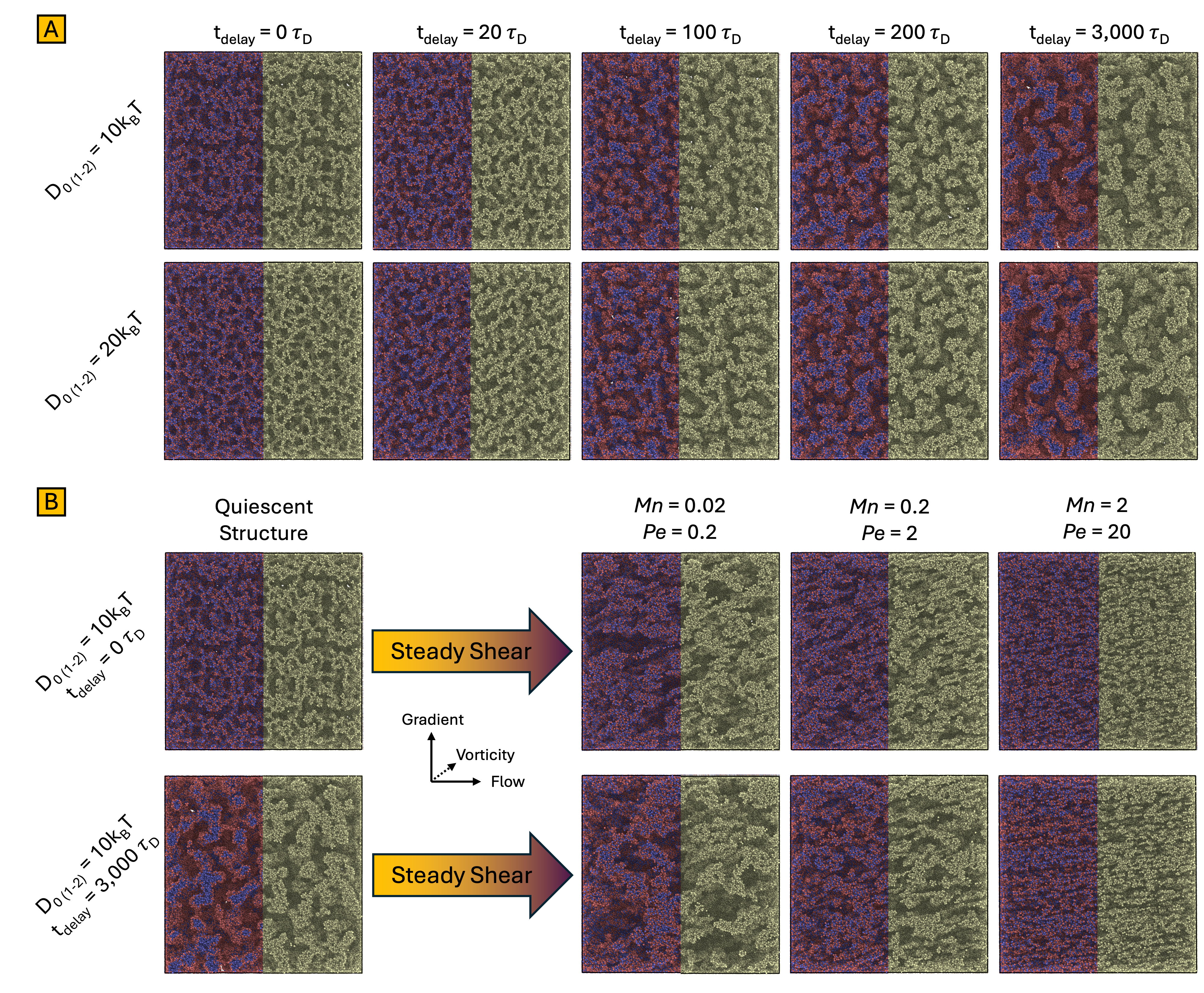}
    \caption{\textbf{[A]} Renditions of quiescent structures constructed with $D_{0(1-2)} = 10k_BT$ (top) and $20k_BT$ (bottom) with various values of $t_{delay}$. \textbf{[B]} Renditions of the final structures of $D_{0(1-2)} = 10k_BT$ gels constructed with $t_{delay} = 0$ (top) and $t_{delay} = 3,000\tau_D$ (bottom) after undergoing 10 strains under a range of shear rates. $C_1$ and $C_2$ particles are colored blue and red, respectively, in the left halves of the renditions, and all particles are colored green in the right halves.}
    \label{fig:Fig2}
\end{figure*}

\begin{figure*}[t]
    \centering
    \includegraphics[width=\textwidth]{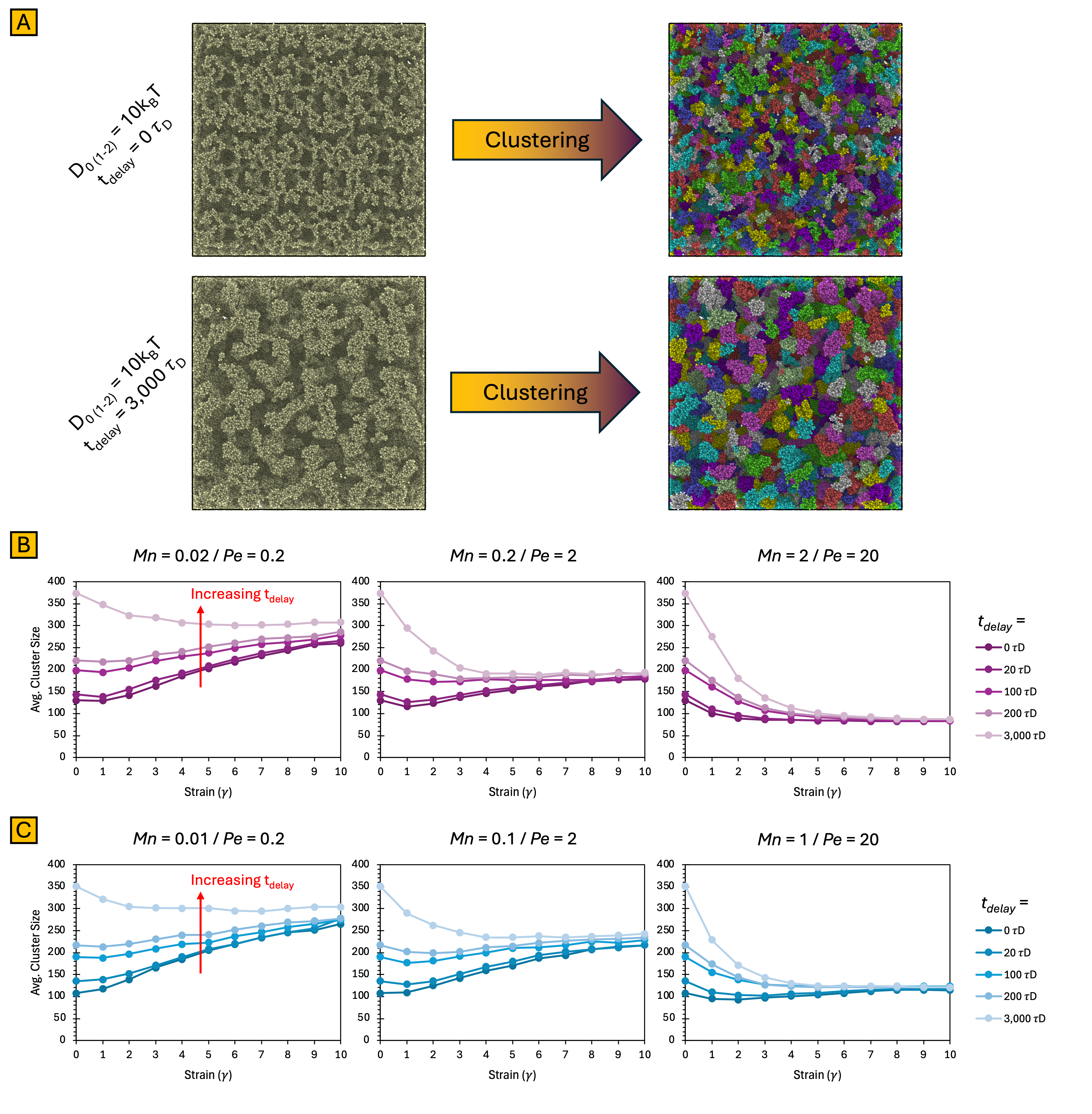}
    \caption{\textbf{[A]} Example of two gels split into clusters via the Infomap algorithm. \textbf{[B-C]} Evolution of the weighted average cluster size of \textbf{[B]} $D_{0(1-2)} = 10k_BT$ binary gels and \textbf{[C]} $D_{0(1-2)} = 20k_BT$ binary gels under steady shear deformation.}
    \label{fig:Fig3}
\end{figure*}

\begin{figure*}
    \centering
    \includegraphics[width=\textwidth]{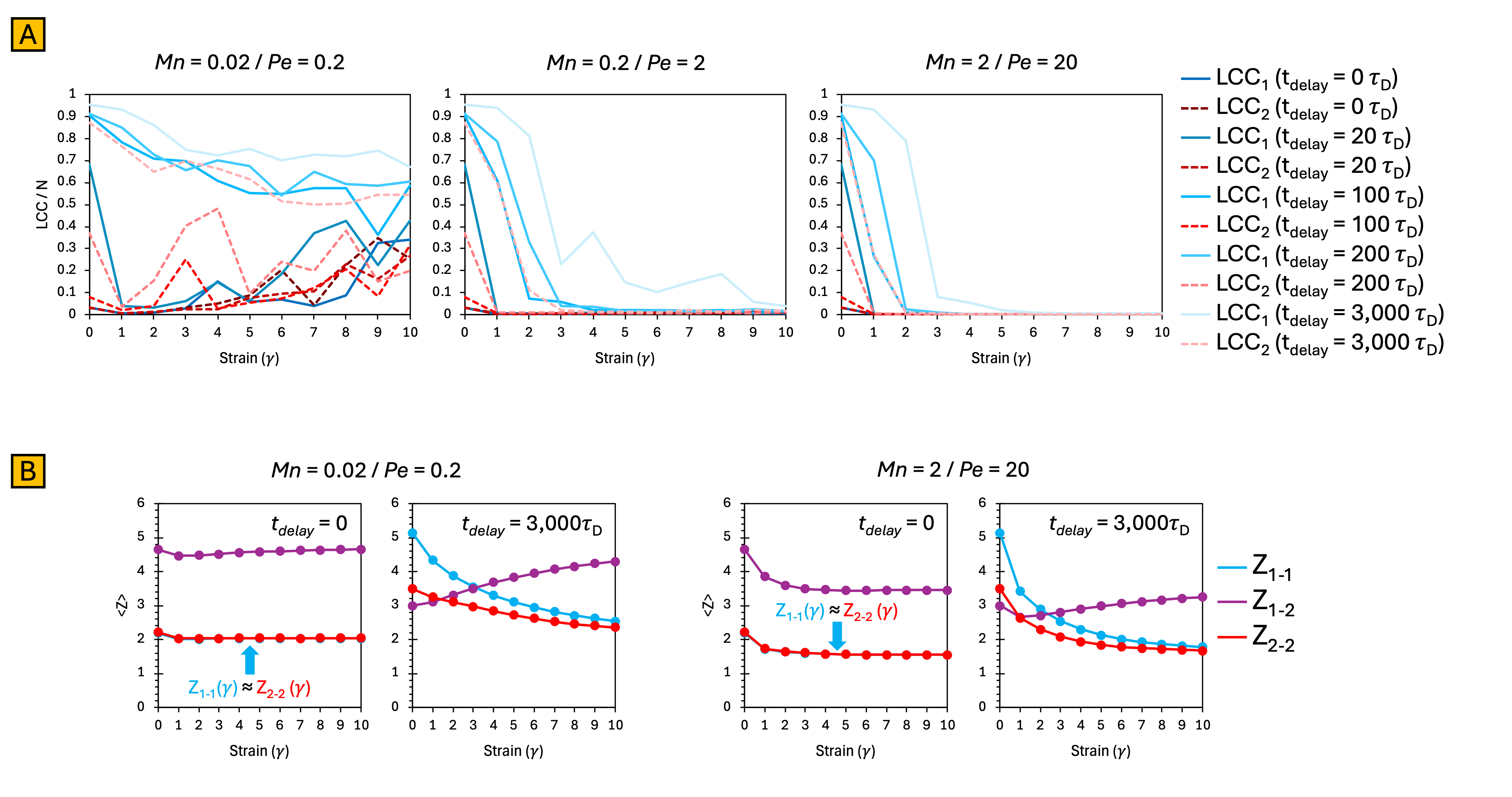}
    \caption{\textbf{[A]} Evolution of single-species largest connected components $LCC_1$ and $LCC_2$ in $D_{0(1-2)} = 10k_BT$ networks, normalized by number of particles $N$. \textbf{[B]} Evolution of species-specific average coordination numbers $Z_{1-1}$, $Z_{1-2}$ and $Z_{2-2}$ of strongly mixed ($t_{delay} = 0$) and de-mixed ($t_{delay} = 3,000\tau_D$) $D_{0(1-2)} = 10k_BT$ gels under slow ($Mn = 0.02 / Pe = 0.2$) and rapid ($Mn = 2 / Pe = 20$) steady shear deformation.}
    \label{fig:Fig4}
\end{figure*}

\begin{figure*}
    \centering
    \includegraphics[width=\textwidth]{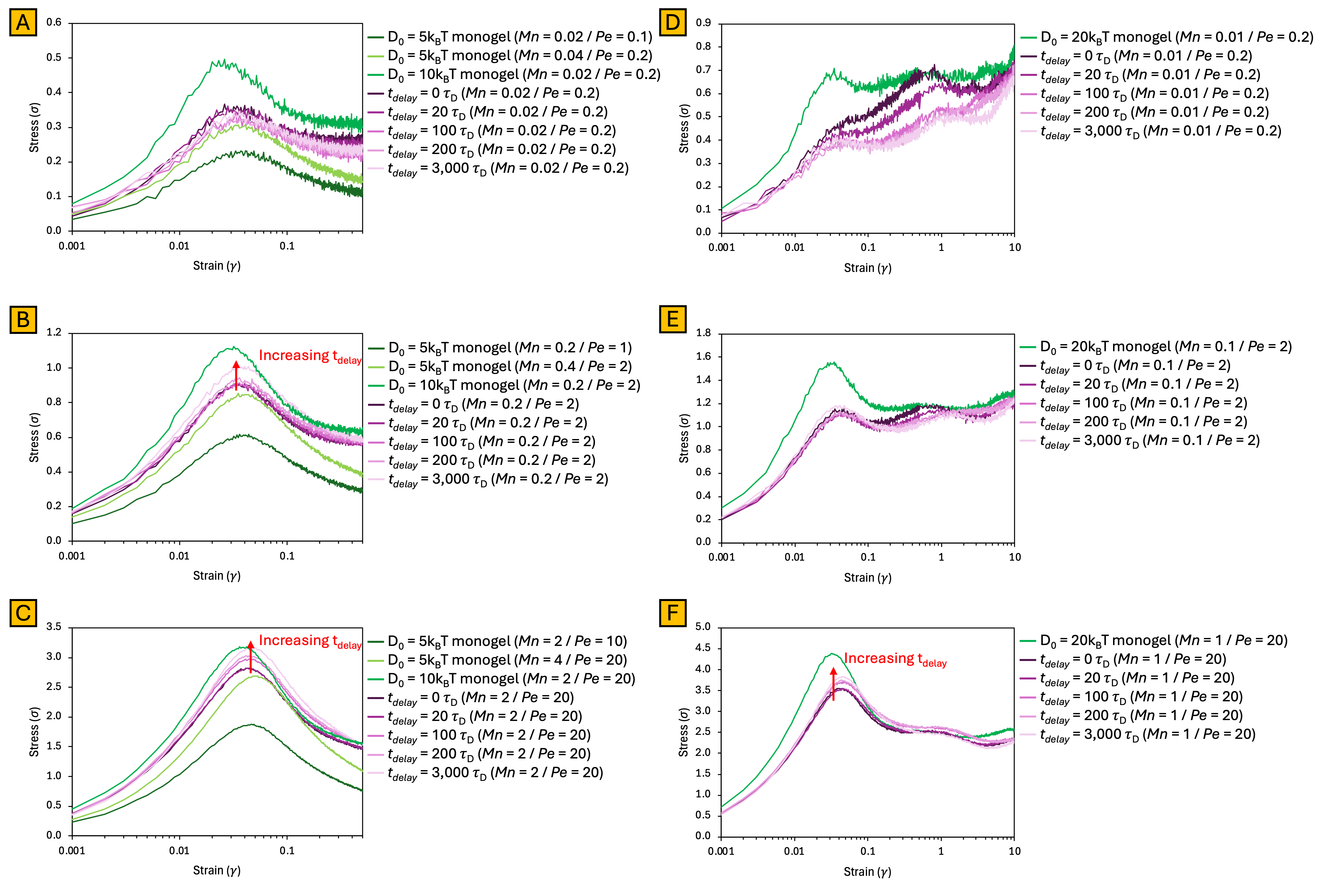}
    \caption{\textbf{[A-C]} Transient stress-strain response of $D_{0(1-2)} = 10k_BT$ binary gels for a range of shear rates, with comparisons to single component gels at the same $Mn$ and/or $Pe$; \textbf{[A]} $Mn = 0.02 / Pe = 0.2$, \textbf{[B]} $Mn = 0.2 / Pe = 2$, \textbf{[C]} $Mn = 2 / Pe = 20$. \textbf{[D-F]} Stress-strain response of $D_{0(1-2)} = 20k_BT$ binary gels for a range of shear rates, with comparisons to single component gels at the same $Mn$ and $Pe$; \textbf{[D]} $Mn = 0.01 / Pe = 0.2$, \textbf{[E]} $Mn = 0.1 / Pe = 2$, \textbf{[F]} $Mn = 1 / Pe = 20$.}
    \label{fig:Fig5}
\end{figure*}

Network science tools have emerged as a compelling way to interrogate mesoscale structure in colloidal and granular materials\cite{Papadopoulos2018}, including methods such as centrality analysis\cite{Mangal2024}, k-core decomposition\cite{Sedes2022}, and various types of community detection algorithms\cite{Huang2016,Bassett2015}. Recent work has also further extended these ideas using probabilistic coarse-graining approaches, such as Gaussian mixture modelling, to infer cluster-level structure in colloidal networks\cite{Nabizadeh2024}.

In this work, we implemented the Infomap algorithm, a high-performance flow-based community detection tool based on the Map Equation\cite{Rosvall2009}, to probe the mesoscale evolution of the studied colloidal networks under shear deformation. This algorithm was used to split the colloidal networks into distinct clusters based on edge lists constructed from neighboring particle pairs (with neighboring particles defined as those whose interparticle surface-surface distance does not exceed 0.1$a$), as demonstrated in Figure \ref{fig:Fig3}A. The weighted average size of these clusters was tracked over the course of 10 applied strains. The Infomap algorithm was selected due to its exceptional computational efficiency and scalability which allows for the practical analysis of the large networks studied here. Analyses with settings utilizing two-level clustering with a  Markov Time of 2 and 100 trials are reported. Here, the Markov Time is a parameter that adjusts the exploration timescale of the random walker used by the algorithm and therefore sets the resolution of the detected clusters\cite{Kheirkhahzadeh2016}. These settings provided sufficient resolution to distinguish differences in the cluster sizes of the quiescent gel structures.

The evolution of the weighted average cluster size under deformation for a range of steady shear conditions is shown in Figure \ref{fig:Fig3}B-C. Under rapid shear deformation ($Mn \geq 1$), the quiescent gel structures undergo a pronounced reduction in average cluster size, consistent with the fluidization of a gel into small flocs expected at such rapid deformations. Under weaker shear deformation ($Mn < 1$), we observe the convergent evolution of the gels towards structures with similar weighted average cluster sizes, despite differences in their quiescent structures arising from variations in $t_{delay}$. Slight variations in weighted average cluster size at 10 strain units only become apparent at the weakest studied shear deformations ($Mn$\textasciitilde $0.01$), with coarser quiescent structures formed with larger $t_{delay}$ displaying slightly larger final cluster sizes. This convergent structural evolution and the weak dependence of final cluster size on $t_{delay}$ were observed over the whole range of tested deformation conditions and for both interaction strengths $D_{0(1-2)} = 10k_BT$ and $20k_BT$. Additional analyses performed using a Markov Time of 3 (Supplementary Material, Figure S2) yield the same overall conclusions, as does visual observation of the sheared structures (Figure \ref{fig:Fig2}B).

\subsection{Largest Connected Component Analysis}

In order to gain insight on species distribution in the emergent structures under deformation, we tracked the evolution of the sizes of the single-species largest connected components ($LCC_1$ and $LCC_2$) over 10 applied strains (Figure \ref{fig:Fig4}A). These are defined as the largest connected networks composed exclusively of particles of a single species (where connectivity is determined by an interparticle surface-surface distance not exceeding $0.1a$). To corroborate and contextualize the trends observed in $LCC$ sizes, we also performed species-specific average coordination number analysis (Figure \ref{fig:Fig4}B). Here, we determine the average number of $C_1$ neighbors surrounding a $C_1$ particle ($Z_{1-1}$), the average number of $C_2$ neighbors surrounding a $C_2$ particle ($Z_{2-2}$) and the average number of $C_2$ neighbors surrounding a $C_1$ particle ($Z_{1-2}$).
At relatively rapid shear deformations ($Mn >> 0.01$), we observe a largely monotonic decrease in the size of both $LCC_1$ and $LCC_2$. We note that for all tested quiescent structures and deformation conditions, the species-indiscriminate $LCC$ accounts for more than $95\%$ of the total colloidal particle population. This indicates that the observed decreases in $LCC_1$ and $LCC_2$ under high $Mn$ deformation arise primarily from increased mixing between colloidal species within the network, rather from than fragmentation of the networks into small, disconnected aggregates. Average coordination number analysis supports this interpretation; after 10 applied strains, $Z_{1-2}$ is consistently higher than $Z_{1-1}$ and $Z_{2-2}$, even in cases where it was substantially lower at quiescent conditions, indicating the emergence of structures dominated by different-species neighbor pairs.

At very weak shear deformations ($Mn$\textasciitilde $0.01$), the same monotonic decreases in $LCC_1$ and $LCC_2$ were not always observed. However, we propose that this is an artifact of the emergence of large particulate aggregates under weak deformation, rather than an absence of species mixing. As the gel structures coarsen under weak deformation, individual particles acquire more neighbors, increasing the likelihood of same-species contacts and consequently, the existence of tenuously connected single-species networks. Average coordination number analysis lends credence to this interpretation, showing that, as before, after 10 applied strains $Z_{1-2}$ is consistently higher than $Z_{1-1}$ and $Z_{2-2}$.

We therefore conclude that, regardless of the applied deformation rate or the extent of species segregation observed in the quiescent structure, the gel networks studied here exhibit a robust tendency towards rearrangement into mixed structures lacking large single-species domains.

\subsection{Stress Response}

When considering the stress response of binary gels, it is useful to draw comparisons to the stress response of single component gels with both weak interparticle interactions ($D_0=5k_BT$) and strong interparticle interactions ($D_0=10k_BT$ when comparing to $D_{0(1-2)} = 10k_BT$ binary gels or $20k_BT$ when comparing to $D_{0(1-2)} = 20k_BT$ binary gels). Single component gels with $D_0  = 5kB_T$, $10k_BT$ and $20k_BT$ exhibit qualitatively similar stress-strain responses, characterized by a single value of the shear stress peak at the overshoot, $\sigma_{max}$, that occurs at similar strain deformation $\gamma_{max}$ for every gel. At a fixed deformation rate, $\sigma_{max}$ increases monotonically with $D_0$, with $D_0  = 20k_BT$ gels exhibiting the highest $\sigma_{max}$, followed by $D_0  = 10k_BT$ and lastly $D_0  = 5k_BT$ gels (Supplementary Material, Figure S3).

Despite the similarity in quiescent structures and trends in structural evolution observed both in systems with $D_{0(1-2)} = 10k_BT$ and with $D_{0(1-2)} = 20k_BT$, variations in the strength of the interspecies interaction $D_{0(1-2)}$ were found to lead to drastically different mechanical responses. In the case of systems with $D_{0(1-2)} = 10k_BT$ (Figure \ref{fig:Fig5} A-C), we observe the emergence of two concurrent trends: First, as shear deformation rate is increased, gels constructed with longer $t_{delay}$ exhibit increasingly higher $\sigma_{max}$ than those with a smaller $t_{delay}$. Secondly, as shear deformation rate is increased, the $\sigma_{max}$ exhibited by the binary gels approached those of single component gels with stronger interparticle attractions ($D_0  = 10k_BT$) rather than ones with weaker interparticle attractions ($D_0  = 5k_BT$). Thus, we observe that while a gel constructed with $D_{0(1-2)} = 10k_BT$ and a large $t_{delay}$ was structurally analogous to a single component gel with weak attractions $D_0  = 5k_BT$\cite{Kaltashov2026}, under rapid shear deformation it also displays mechanical properties similar to those of a single component gel with stronger attractions $D_0  = 10k_BT$.

In the case of $D_{0(1-2)} = 20k_BT$ systems (Figure \ref{fig:Fig5} D-F), we observe a more complex stress response. At rapid shear deformation ($Mn >> 0.1$), the behavior qualitatively mimics that of $D_{0(1-2)} = 10k_BT$ systems, exhibiting a single $\sigma_{max}$ that increases monotonically with $t_{delay}$. At lower shear deformation rates ($Mn << 1$), however, the stress-strain response displays two-step yielding, characterized by the emergence of two distinct stress peaks. At the weakest applied shear deformations ($Mn$\textasciitilde $0.01$), we further observe that gels constructed with lower $t_{delay}$ consistently exhibit lower $\sigma$ than those constructed with larger $t_{delay}$. Interestingly, in this low $Mn$ regime, the two-step yielding behavior disappears in gels constructed without sequential gelation, indicating that structural heterogeneity is essential for the emergence of multiple yielding events under slow deformation.

\section{Discussion and Conclusions}

The central question addressed in this study is whether the formation pathway of a multicomponent colloidal gel remains mechanically relevant once the material is subjected to steady shear. Specifically, we asked whether binary gels formed through distinct sequential-gelation histories become rheologically equivalent if shear drives them toward structurally similar states. Our results show that they do not. Although dominant interspecies attractions promote shear-induced convergence in several network-scale structural descriptors, the transient stress response remains strongly dependent on both gelation delay and interspecies attraction strength. This decoupling between apparent structural convergence and persistent rheological memory highlights the importance of formation history in determining the flow response of multicomponent colloidal gels.

These results motivate a distinction between morphological memory and mechanical memory in sequentially gelled multicomponent networks. Morphological memory refers to the persistence of visible or connectivity-based signatures of the initial assembly pathway, such as domain size, single-species connectivity, or degree of mixing. Mechanical memory, by contrast, refers to the persistence of pathway-dependent deformation and yielding behavior even after such structural signatures have been substantially reduced. In the present systems, steady shear largely suppresses morphological memory according to the structural descriptors considered here, but the transient stress response retains a clear mechanical memory of the gelation delay. This distinction is important because it implies that post-shear structure alone may be insufficient to infer the yielding history or flow response of multicomponent colloidal gels.

Previously, we showed that binary gels with strong interspecies and weak intraspecies attractions can be programmed through the sequential gelation delay, with increasing delay producing a transition from fine, well-mixed networks to coarser, more segregated shell-core-like morphologies\cite{Kaltashov2026}. Here, we find that these initially distinct quiescent structures undergo substantial restructuring under steady shear. Across the deformation conditions studied, the networks tend to rearrange toward more mixed configurations dominated by interspecies contacts. The characteristic feature sizes of the sheared structures are governed primarily by the imposed shear rate rather than by the initial quiescent morphology. Thus, gels that originate from markedly different formation histories can evolve toward comparable structural states when subjected to the same shear conditions (Figure \ref{fig:Fig2}B and Supplementary Material, Videos S1-6). However, the convergence observed here should be interpreted specifically as convergence in the connectivity-based descriptors considered in this study, rather than complete structural equivalence. Metrics such as average cluster size, largest connected components, and coordination number quantify the extent of mixing and network connectivity, but they do not distinguish between contacts that are mechanically load bearing and contacts that are geometrically present but weakly stressed. Similarly, these descriptors do not encode bond age, bond orientation, anisotropic force transmission, or the sequence by which contacts are broken and reformed during deformation. Two networks may therefore converge toward similar values of common scalar structural metrics while retaining distinct stress-bearing backbones and deformation pathways. This distinction is likely central to the persistence of formation-history-dependent yielding observed here.

However, this structural convergence does not translate into rheological equivalence. For systems with $D_{0(1-2)} = 10 k_BT$, the stress overshoot at a fixed shear rate increases with increasing gelation delay, and this dependence becomes most pronounced at high shear rates. This result is notable because gels formed with long delays possess coarse quiescent morphologies that resemble weakly attracted networks, yet under rapid deformation they exhibit stress responses closer to those of strongly attracted single-component gels. In these systems, the initial architecture influences how force is transmitted and released during deformation, even after shear has substantially reorganized the network structure. The comparison with single-component gels further emphasizes that quiescent morphology alone is insufficient to predict the transient rheology of the binary networks. Long-delay binary gels possess coarse morphologies that resemble weakly attracted single-component gels, but under rapid deformation their stress overshoots approach those of more strongly attracted single-component gels. This behavior suggests that rapid shear activates strong interspecies contacts embedded within the heterogeneous binary architecture. Thus, the relevant mechanical unit under deformation is not simply the apparent domain morphology, but the subset of strong contacts that become engaged during loading.

The response of the $D_{0(1-2)} = 20 k_BT$ systems reveals an additional layer of complexity. Although their quiescent morphologies are broadly similar to those of the $D_{0(1-2)} = 10 k_BT$ systems, their stress response is qualitatively different, particularly at low shear rates, where two-step yielding emerges. This behavior suggests that slow deformation resolves two distinct structural relaxation processes. We hypothesize that the first stress maximum reflects deformation and partial collapse of initially segregated, weakly connected domains, whereas the second maximum reflects subsequent loading and rearrangement of a mixed network increasingly dominated by strong $C_1-C_2$ contacts. At higher deformation rates, these processes are compressed into a narrower strain window and appear as a single dominant stress overshoot. Thus, the disappearance of two-step yielding at high rate does not necessarily imply the absence of multiple microscopic processes, but rather that the imposed deformation does not allow them to appear as distinct macroscopic events. The absence of two-step yielding in the non-sequentially gelled system supports this interpretation: when the initial structure is already well mixed, shear does not encounter a separate mesoscale reorganization process before loading the interspecies network.

A notable feature of the present results is that increasing the interspecies attraction from $10 k_BT$ to $20 k_BT$ does not simply amplify the stress response; it changes its qualitative form. Although the two interaction strengths produce broadly similar quiescent morphologies and both undergo shear-induced mixing, only the stronger interspecies attraction produces pronounced two-step yielding at low deformation rates. This indicates that interspecies attraction affects not only the magnitude of the stress response, but also the sequence of deformation mechanisms available to the network. In this sense, interspecies attraction acts as a control parameter that determines whether shear-induced restructuring occurs smoothly or through mechanically distinct yielding events.

These findings suggest that the mechanical response of sequentially gelled binary colloidal networks is controlled by an interplay among initial mesoscale heterogeneity, interspecies bond strength, and deformation rate. Strong interspecies attractions drive the system toward mixed configurations under shear, but the pathway by which that mixed state is reached depends on the initial arrangement of domains and contacts. As a result, common structural descriptors measured after deformation may not be sufficient to predict the transient stress response. Instead, the stress-strain behavior retains a measurable memory of the assembly pathway. From a materials-design perspective, these results show that formation sequence can serve as an independent control parameter for programming transient rheology. In conventional design of colloidal gels, yielding is often tuned by changing volume fraction, attraction strength, particle size, or composition. Here, the composition and interaction parameters are fixed within each family of simulations, yet changing the gelation delay substantially modifies the stress overshoot and, in strongly interacting systems, determines whether yielding occurs through one or two apparent steps. Sequential gelation therefore provides a route to tune mechanical response through processing history rather than formulation alone.

A limitation of the present study is the use of Brownian Dynamics during shear deformation. The large system size considered here, $N=442,206$ particles, was chosen to avoid artificial confinement of mesoscale structural features and to improve the statistical reliability of rheological measurements. This size, however, makes Brownian Dynamics an efficient and practical simulation method. Its efficiency comes at the cost of over simplifying hydrodynamic interactions, which are known to influence the shear-induced restructuring of colloidal gels, including phenomena such as vorticity-aligned density fluctuations\cite{Varga2018}. The absence of hydrodynamic interactions may explain why the sheared structures remain tenuously connected even at high deformation rates, with the species-indiscriminate largest connected component continuing to contain majority of the particles. While hydrodynamic interactions may alter the absolute degree of shear-induced densification, fragmentation, or vorticity alignment, the central conclusion of this study concerns relative comparisons across gels with different formation histories subjected to the same deformation protocol. The persistence of delay-dependent stress responses despite convergence in connectivity-based descriptors may therefore be more robust than the detailed morphology of the sheared state. Future work incorporating hydrodynamic interactions, for example through Brownian Dynamics with a Rotne-Prager-Yamakawa mobility tensor or related methods, would be valuable for testing the robustness of the trends observed here.

Experimentally, the present findings suggest that two multicomponent gels with similar post-shear microstructures may nevertheless display different start-up rheology if their assembly pathways differ. This has implications for materials processed through mixing, pH jumps, thermal ramps, salt addition, or staged aggregation protocols, where different components may become attractive at different times. In such systems, rheological characterization after pre-shear may not fully erase the influence of formation history, especially when strong interspecies attractions create persistent load-bearing contacts. Taken together, these results show that sequential gelation provides a route to encode rheological memory into multicomponent colloidal materials. Even when start-up shear reduces differences in common structural descriptors, the transient stress response can retain a measurable signature of the assembly pathway. This finding suggests that formation sequence should be treated as an independent design variable, alongside interaction strength and composition, in the engineering of multicomponent colloidal gels with programmable yielding and flow behavior.

\section{Supplementary Material}
The supplementary material contains (Section S1) a description of the numerical methods used to model quiescent structure self-assembly; (Figure S1) the solvent-accessible surface areas of the quiescent structures; (Figure S2) evolution of the weighted average cluster size of $D_{0(1-2)}=10k_BT$ networks from two-level Infomap analysis (Markov time 3); (Figure S3) stress–strain responses of single-component gel benchmarks; (Videos S1-6) visual renditions of $D_{0(1-2)}=10k_BT$ networks subjected to steady shear deformation.

\begin{acknowledgments}
This work was supported by the U.S. Naval Research Laboratory (Grant No. N000142312772). Computational resources were provided by the Massachusetts Green High-Performance Computing Center in Holyoke, MA.
\end{acknowledgments}


\section*{Data Availability Statement}
The data that support the findings of this study are openly available in https://github.com/procf

\section*{References}

\nocite{*}
\bibliography{aipsamp}

\end{document}